\documentclass[conference]{IEEEtran}
\IEEEoverridecommandlockouts

\usepackage[english]{babel} 
\usepackage[T1]{fontenc} 
\usepackage[utf8]{inputenc} 
\usepackage{graphicx}

\usepackage[hyphens]{url}
\usepackage[nospace,noadjust]{cite}
\usepackage{amsmath,amsfonts,amsthm}
\usepackage[caption=false,font=footnotesize,hangindent=10pt]{subfig}
\usepackage[linesnumbered,vlined,boxed,ruled]{algorithm2e}
\usepackage{color,xparse}
\usepackage{tabu}
\usepackage{hyperref}

\usepackage{multirow}
\usepackage{enumitem}
\usepackage{booktabs}

\usepackage{caption}
\usepackage[capitalise]{cleveref}
\usepackage{bbding}
\usepackage{makecell}
\usepackage{wrapfig}

\usepackage{titlesec}
\titlespacing\section{0pt}{4pt}{0pt}
\titlespacing\subsection{0pt}{4pt}{0pt}
\titlespacing\subsubsection{0pt}{4pt}{0pt}

\graphicspath{{./}{figure/}}

\SetVlineSkip{2pt}

\SetCommentSty{myAlgComment}
\SetKwProg{Fn}{Function}{:}{}
\SetAlFnt{\small}
\SetAlCapFnt{\small}
\SetAlCapNameFnt{\small}

\newcommand{\header}[1]{\noindent\textbf{#1}}

\newtheorem{problem}{Problem}

\newcommand{\Real}{\mathbb{R}}
\newcommand{\CI}{\mathcal{I}}
\newcommand{\CA}{\mathcal{A}}
\newcommand{\CC}{\mathcal{C}}
\newcommand{\CD}{\mathcal{D}}
\newcommand{\CM}{\mathcal{M}}
\newcommand{\CF}{\mathcal{F}}
\newcommand{\CK}{\mathcal{K}}
\newcommand{\CQ}{\mathcal{Q}}
\newcommand{\CT}{\mathcal{T}}
\newcommand{\CP}{\mathcal{P}}

\title{\fontsize{18pt}{22pt}\selectfont BSODiag: A Global Diagnosis Framework for Batch Servers Outage\\ in
  Large-scale Cloud Infrastructure Systems\vspace{-0.6cm}}

\author{%
  \IEEEauthorblockN{Tao Duan\IEEEauthorrefmark{2},
                    Runqing Chen\IEEEauthorrefmark{3},
                    Pinghui Wang\IEEEauthorrefmark{2}$^*$\thanks{* Corresponding Author},
                    Junzhou Zhao\IEEEauthorrefmark{2}$^*$,
                    Jiongzhou Liu\IEEEauthorrefmark{3},
                    Shujie Han\IEEEauthorrefmark{3},
                    Yi Liu\IEEEauthorrefmark{3} and Fan Xu\IEEEauthorrefmark{3}}
  
  \IEEEauthorblockA{\IEEEauthorrefmark{2}{\em MOE KLINNS Lab, Xi'an Jiaotong University, Xi'an 710049, P.~R.~China}\\
                    \IEEEauthorrefmark{3}{\em Alibaba Cloud Computing, Hangzhou 310030, P.~R.~China}
}
 \IEEEauthorblockA{
                    duantao@stu.xjtu.edu.cn, \{phwang, junzhou.zhao\}@xjtu.edu.cn,
      \\ \{runqing.rq, jiongzhou.ljz, mars.ly,xufan.xf\}@alibaba-inc.com, shujiehan00001@gmail.com
 \vspace{-0.4cm}
 }
}

\begin{document}

\maketitle

\begin{abstract}
Cloud infrastructure is the collective term for all physical devices within cloud systems.
Failures within the cloud infrastructure system can severely
compromise the stability and availability of cloud services. Particularly, batch servers outage,
which is the most fatal failure, could result in the complete unavailability of all
upstream services. In this work, we focus on the batch servers outage diagnosis problem,
aiming to accurately and promptly analyze the root cause of outages to facilitate troubleshooting.
However, our empirical study conducted in a real industrial system indicates that it is a challenging task. 
Firstly, the collected single-modal coarse-grained failure monitoring data (i.e., alert, incident, or change)
in the cloud infrastructure system is insufficient for a comprehensive failure profiling. Secondly,
due to the intricate dependencies among devices, outages are often the cumulative result of multiple failures,
but correlations between failures are difficult to ascertain.
To address these problems, we propose BSODiag, an unsupervised and lightweight diagnosis framework for batch servers outage. BSODiag provides a global analytical perspective, thoroughly explores
failure information from multi-source monitoring data, models the spatio-temporal correlations among failures,
and delivers accurate and interpretable diagnostic results.
Experiments conducted on the Alibaba Cloud infrastructure system show that BSODiag achieves
87.5\% PR@3 and 46.3\% PCR, outperforming baseline methods by 10.2\% and 3.7\%, respectively.

\end{abstract}

\section{Introduction}
\label{sec:introduction}

Cloud systems have gained great popularity in a variety of applications such as
computing, storage, and e-commerce, owing to their superiority in deployment,
migration, and scalability, when compared to traditional
systems~\cite{chen2019outage}.
Cloud service providers (CSPs) like AWS, Azure, and Alibaba Cloud, offer both private
and public cloud services to their clientele by leasing cloud devices or selling
cloud products~\cite{cloud2024}.
Cloud infrastructure, constructed from essential physical equipment such as
computing units, networking devices, and power components, forms the backbone of
the cloud system.
Due to its vast system scale and intricate architecture, failures and faults
within the cloud infrastructure system severely undermine the stability and
availability of upstream services, potentially leading to customer attrition and
revenue loss.
It is thus an important task to diagnose failures and faults within the cloud
infrastructure system promptly and accurately.

In a cloud infrastructure system, a {\em batch servers outage}, i.e., the
simultaneous breakdown of a cluster of related servers, is often considered as
one of the most fatal failures.
Such a failure could typically result in the complete unavailability of
all upstream services.
For instance, an actual batch servers outage occurred
at the Alibaba Cloud Hong Kong data center on December 18, 2022, causing a
catastrophic interruption in elastic container service (ECS), PolarDB, and networking
service, lasting for over $12$ hours~\cite{outage2022}.

\begin{figure}[t]
  \centering
  \includegraphics[width=.8\linewidth]{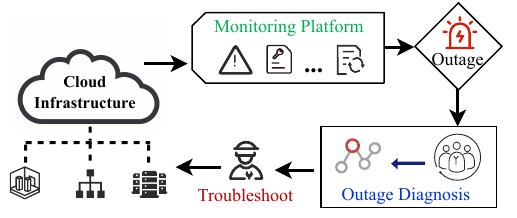}
  \caption{The life cycle of a batch servers outage diagnosis.
  }
  \label{fig:cloud}
  \vspace{-0.5cm}
\end{figure}

A cloud infrastructure system is a large-scale system comprising thousands to
millions of devices.
In practice, the devices within a cloud infrastructure system are usually
grouped into several domains according to their physical functionalities, e.g.,
Internet data center (IDC), cloud networking, and cloud servers, to facilitate
the dispatching and diagnosis of failures~\cite{wang2021outage}.
\cref{fig:cloud} shows the typical life cycle of a batch servers outage diagnosis.
An online monitoring platform tracks and records the cloud infrastructure system
status continuously.
When a batch servers outage event is reported, experts from different domains are
dispatched to meticulously analyze the monitoring data and diagnose the cause of
the batch servers outage.
On-site engineers (OSEs) then use these diagnostic results to troubleshoot and
restore system availability.
In this process, the unique domain-specific failure knowledge from different
expert teams is crucial for an accurate diagnosis.
Despite the widespread use of this standardized diagnosis workflow, our
observations from real-world industrial practice reveal that, since it entails
manually analyzing the monitoring data and intensive collaboration among
different specialties, this makes diagnosis time-consuming and labor-intensive.
This substantially hampers the ability to respond quickly to failures.

In recent years, many efforts have been devoted to providing an automatic
failure diagnosis strategy for microservice
systems~\cite{ma2020diagnosing,ikram2022root,ma2021servicerank,li2022actionable}.
As small-scale systems that typically only involve at most dozens of
components, microservice systems extensively collect fine-grained machine data
such as metrics~\cite{zhang2022metric}, logs~\cite{meng2019loganomaly}, and
traces~\cite{liu2020trace} to diagnose failures and model the correlations
between failures~\cite{chen2023tele}.
In contrast, in a large-scale cloud infrastructure system, the monitoring
platform is constrained by storage and computing resources, allowing only the
recording of coarse-grained monitoring data such as alerts~\cite{zhao2020understanding},
incidents~\cite{chen2020towards}, and changes~\cite{zhao2023identifying}.
Significant disparities in data formats make the existing methods tailored for
microservice systems entirely inapplicable to our batch servers outage diagnosis
problem.
Some recent methods attempt to address the large-scale system
failure diagnosis problem through failure graph
analysis~\cite{wang2021outage,chakraborty2023esro,wang2021groot}.
However, due to the inability to accurately measure the correlation between
failures, these methods suffer from the underreporting and misreporting issues.
This motivates us to propose a novel solution.

In this work, we study the problem of {\em batch servers outage diagnosis in
  large-scale cloud infrastructure systems}, where the goal is to accurately and
promptly locate the root cause of an outage, and facilitate troubleshooting.
Our in-depth empirical study conducted on Alibaba Cloud reveals pivotal
challenges to solve this problem.
First, coarse-grained monitoring data collected in the cloud infrastructure
system only provides a cursory aggregation for suspicious anomalies and system
status changes, rendering them insufficient for comprehensive failure profiling.
Second, the correlations among various failures vary significantly across
domains.
For example, network failures often exhibit a hierarchical structure, while IDC
failures often propagate through resource sharing among devices.
This makes it difficult to provide an accurate global measure of failure
correlation.
Third, due to the intricate dependencies among devices, outage is frequently
the outcome of the cumulative effect of multiple failures.
This not only makes it challenging to accurately identify the root cause, but
also results in locating a single root cause that fails to provide OSEs with a
holistic perspective of the failure propagation.
Therefore, an accurate and interpretable failure root cause analysis method is
desired.

Motivated by these observations, we propose BSODiag, an {\em unsupervised} and
{\em lightweight} framework designed for diagnosing a batch servers outage.
The core ideas of BSODiag are: (i) fully exploiting multi-source monitoring data
via modality-specific failure detection and a unified failure event
representation to provide a more comprehensive failure profiling; (ii)
synergistically examining the spatio-temporal information in failures, utilizing
the failure knowledge contained in historical outages and the device
dependencies reflected in current outage, to accurately quantify the correlation
between failures.
Based on this global analytical perspective, BSODiag combines the two diagnostic
tasks, i.e., root cause localization and failure propagation path inference, to
enhance the accuracy and interpretability of diagnosis results.
Moreover, the lightweight design of BSODiag ensures a rapid
response to outages.

In more detail, BSODiag comprises three pivotal modules to tackle the
aforementioned challenges individually.
First, a {\em multi-source failure detection} module employs modality-specific
components to simultaneously detect outage-related failures from alerts,
incidents, and changes.
Alerts are converted into time series through a serialization strategy, enabling
unsupervised anomaly detection methods to pinpoint authentic failures.
Changes are filtered using an expert-provided whitelist, and only high-risk
actions are retained.
The related failures are uniformly integrated into events, and further represented as attribute tuples.
Subsequently, a {\em failure correlation mining} module combines failure
knowledge mining and knowledge verification to discern the failure correlations
reflected in historical data, and integrate them into a failure knowledge graph.
Finally, an {\em outage root cause analysis} module constructs an event cause
graph using the failure knowledge graph and a device dependency graph, and delivers
interpretable diagnostic results through a customized random walk and a propagation
probability-based path inference.

Our contributions can be summarized as follows:
\begin{itemize}
\item To the best of our knowledge, we are the first to study the batch servers
  outage diagnosis problem, which demands urgent resolution in the industry.
  We conducted an empirical study on a large-scale cloud system, and the
  observations uncover the key insights of this problem.

\item We propose BSODiag, an unsupervised and lightweight batch servers outage
  diagnosis framework to address the problem.
  BSODiag thoroughly explores failure information from multi-source monitoring
  data, models the spatio-temporal correlations among failures, and delivers
  accurate and interpretable diagnosis results.

\item To evaluate the effectiveness of BSODiag, we collected real-world failure
  monitoring data from Alibaba Cloud infrastructure system.
  Our extensive experiments conducted on this industrial dataset show
  that BSODiag achieves $87.5\%$ on PR@3 and $46.3\%$ on PCR, outperforming
  baseline methods by $10.2\%$ and $3.7\%$, respectively.
\end{itemize}

\section{Background and Empirical Observations}
\label{sec:background}

In this section, we first describe related background about the cloud
infrastructure system and the collected failure monitoring data to facilitate a
comprehensive understanding of the proposed batch servers outage diagnosis
problem.
Then, we conduct an in-depth empirical study in a real-world industrial context
to elicit crucial insights for addressing this problem.

\begin{figure*}[htp]
  \centering
  \subfloat[An Alert Sequence \label{fig:alert}]{%
    \includegraphics[width=0.32\linewidth]{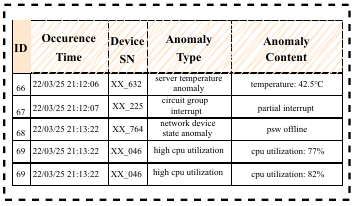}}
  \subfloat[A Batch Servers Outage Incident \label{fig:incident}]{%
    \includegraphics[width=0.32\linewidth]{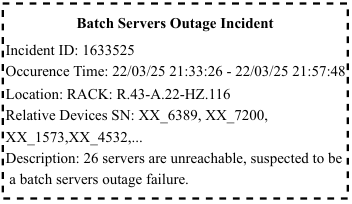}}
  \subfloat[A Refrigerant Replace Change \label{fig:change}]{%
    \includegraphics[width=0.32\linewidth]{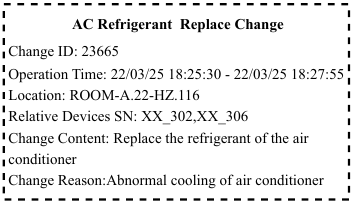}}
  \caption{Different failure monitoring data collected in cloud
    infrastructure system.\label{fig:data}}
\vspace{-0.5cm}
\end{figure*}

\subsection{Cloud Infrastructure System}

A cloud infrastructure system (CIS) refers to the collective term for all
physical devices utilized in a cloud system.
In current industrial practice, devices in a cloud infrastructure system are
categorized into distinct domains based on their physical functionalities,
maintained by experts with domain-specific knowledge to facilitate failure
management and diagnosis.
The Alibaba Cloud infrastructure system, as discussed in this paper, is grouped
into three primary domains: Internet data center, cloud networking, and cloud
servers.

\header{Internet Data Center (IDC)}.
The primary role of an IDC is to maintain standardized rooms to ensure the
operation of servers and network devices.
It primarily comprises an uninterrupted power supply (UPS) system and a temperature
and humidity control system.
Owing to the inherent physical propagation mechanisms, a failure within the IDC
domain commonly results in extensive batch failures~\cite{roy2017passive}.
For instance, a power failure in a server rack can cause all servers on that
rack to go offline, and a temperature control system failure can result in
malfunctioning of all the devices in this room.

\header{Cloud Networking}.
Cloud networking devices, such as switches, and load balancers, provide
connectivity among servers, and between servers and the backbone network.
Owing to hierarchical structure and redundant design, networking devices
exhibit complex interdependencies, causing failures within the network domain
to manifest significant cascading effects.
For example, a congestion failure in a top-layer load balancer can disrupt the
operation of mid-layer switches, and ultimately cause bottom-layer servers to go offline.

\header{Cloud Servers}.
Cloud servers are essential devices that provide computing and storage
capabilities within CIS.
It comprises numerous individual servers which are centrally deployed and
configured into a cluster using virtualization and resource management
techniques.
Each server is equipped with computing units (e.g., CPU and GPU), storage units
(e.g., SSD), and power supply components.
Notably, the failure occurring within each server's component only
impacts the status of the corresponding single server and does not cause a batch
servers outage failure.
Therefore, within the server domain, our focus is solely on batch failures
related to outages.

\subsection{Coarse-grained Failure Monitoring Data}

To enable the timely detection and diagnosis of failures, cloud infrastructure
systems are equipped with an automated monitoring platform~\cite{li2021automated} that continuously
collects failure monitoring data, such as alerts, incidents, and changes, to
document all anomalies, failures, and configuration changes.
In \cref{fig:data}, we present the different types of failure monitoring data collected
in the Alibaba CIS.

\header{Alert}.
Alert data is the structured table that chronologically records anomalies
detected in a CIS.
Each alert, i.e., each row in \cref{fig:alert}, comprises a timestamp, device
serial number, anomaly type, and a concise anomaly description.
It is important to note that an alert simply reports an anomaly and does not
equate to a failure.
For example, an alert for ``high CPU utilization'' may be triggered by server
CPU utilization exceeding a predetermined threshold since increased service
traffic, which does not necessarily imply an actual CPU failure.
This threshold-based triggering mechanism makes alerts sensitive to minor
anomalies but also prone to significant false positives and alert
flooding~\cite{zhao2020understanding,chen2019outage}.

\header{Incident}.
An incident is a textual ticket that aggregates relevant failures into an event.
Each incident details the failure's start time, end time, device serial numbers,
failure type, and a description.
Incidents serve as the most direct evidence for experts to comprehend and
diagnose failures.
However, their intrinsic aggregation mechanism results in two notable
shortcomings, i.e., delayed reporting and omissions.

\header{Change}.
A change records a configuration change event in the CIS, such as manual hardware
replacement, software update, or automated migration of virtual resources.
Industry experiences have shown that changes are commonly high-risk factors
leading to system failures~\cite{zhao2023identifying}.
Following a failure, the change data can assist engineers in checking for
configuration errors and performing rapid rollbacks.

\subsection{Empirical Observations}
\label{sec:empirical_analysis}

In this section, we conduct an in-depth empirical analysis on Alibaba CIS to
answer the following research questions:
\begin{itemize}
\item \textbf{RQ1}: Can coarse-grained monitoring data collected in a CIS adequately describe
  all failures, if not, how to obtain a more comprehensive failure profiling?
\item \textbf{RQ2}: What is the cause of batch servers outage, and what is the
  correlation mechanism between failures?
\item \textbf{RQ3}: What are the necessary diagnostic results for real-world
  applications?
\end{itemize}

\header{Analysis of Data Quality (RQ1)}.
Constrained by storage and computational resources, large-scale cloud
infrastructure systems collect only coarse-grained monitoring data
to diagnose system failures.
Existing methods usually analyze alert, incident, and change data independently
to pinpoint the root cause.
To evaluate the representational efficacy of these monitoring data for all failures,
we collected real-world data from Alibaba CIS and conduct a qualitative
analysis.
Specifically, we select a batch servers outage case and trace six types of
failures occurred near to the outage (i.e., within two hours before the outage and $15$ minutes after the outage).

\begin{table}[htp]
  \centering
  \caption{Analysis of monitoring data quality. 
  }
  \label{tab:data_analysis}
  \begin{tabular}{c|cccc}
    \hline
    \hline
    Failure Type          & Incident   & Change     & Alert      & \#Failures \\
    \hline
    Switch Reboot         & \Checkmark &            &            & 4          \\
    Temperature Anomaly   & \Checkmark &            & \Checkmark & 126        \\
    Refrigerant Replacing &            & \Checkmark &            & 1          \\
    PSU Power Outage      & \Checkmark &            &            & 2          \\
    High CPU Utilization  &            &            & \Checkmark & 305        \\
    Partial Network Loss  &            &            & \Checkmark & 206        \\
    \hline
  \end{tabular}
  \vspace{-0.2cm}
\end{table}

We observe that single-source monitoring data are insufficient to reveal all
suspicious failures.
For instance, incidents are adept at detecting significant failures, such as
``Switch Reboot'' and ``PSU Power Outage'', yet they may overlook minor failures
like ``High CPU Utilization''.
In contrast, alarms are sensitive to these minor failures, but their
``sensitivity" also leads to a large number of false positives regarding
``Partial Network Loss''.
Hence, for a comprehensive and accurate detection of all potential
failures, {\em synchronous analysis of multi-source monitoring data is
  imperative}.

\header{Analysis of Failure Correlation (RQ2)}.
In a CIS, intricate interdependencies exist among devices.
A failure can trigger a cascade of related failures on adjacent devices.
These failures gradually propagate throughout the cloud infrastructure,
ultimately becoming potential factors leading to batch servers outage.
When a batch servers outage failure is detected, the diagnosis system is
expected to accurately identify the root cause of these outage-related failures.
In this process, accurately measuring the correlation among failures is
essential.
To gain a comprehensive understanding of the outages and related failures, we
conducted an analysis of the root cause distribution and the failure correlation
patterns.

\begin{figure}[t]
  \centering
  \vspace{-0.5cm}
  \subfloat[Root Causes Distribution\label{fig:root_cause}]{%
    \includegraphics[width=.4\linewidth]{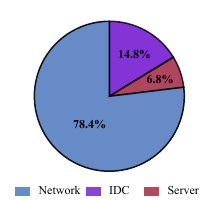}}
  \quad
  \subfloat[Failure Correlation Patterns\label{fig:correlation}]{%
    \includegraphics[width=.4\linewidth]{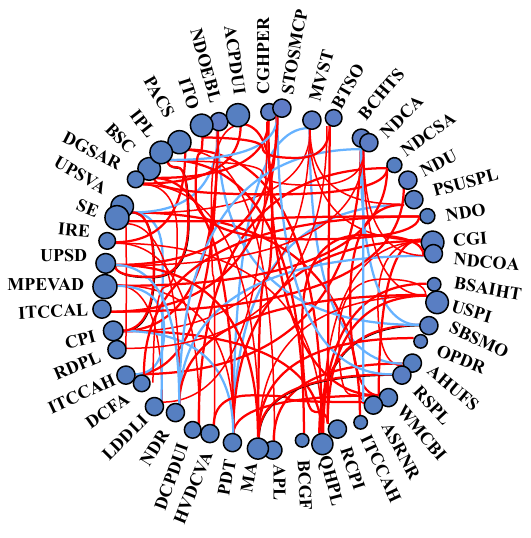}}
  \caption{Analysis of root causes and failure
    correlations.\label{fig:failure_analysis}}
    \vspace{-0.3cm}
\end{figure}

We built a large-scale testing platform in Alibaba CIS for outage data collection
and verification. Figure~\ref{fig:root_cause} depicts the distribution of root
causes of $95$ outage cases that were collected on this testing platform.
Our observation reveals that cross-domain network failures (accounting for
$78.4\%$) and IDC failures (accounting for $14.8\%$) are the primary root causes
of batch servers outage.
Furthermore, \cref{fig:correlation} illustrates the correlation patterns among
$48$ failures selected at random, where two connected nodes represent two
related failures.
Our analysis indicates that failure correlations are intricate, encompassing
both intra-domain (blue connection lines) and cross-domain (red connection lines) correlations.
These findings imply that a batch servers outage often results from concurrent
multi-domains failures.
Hence, {\em it is crucial to develop a failure correlation measurement
technique that can model failure correlations from a global
  perspective} for accurately tracing the root causes of outage.

\begin{figure}[t]
  \centering
  \includegraphics[width=0.7\linewidth]{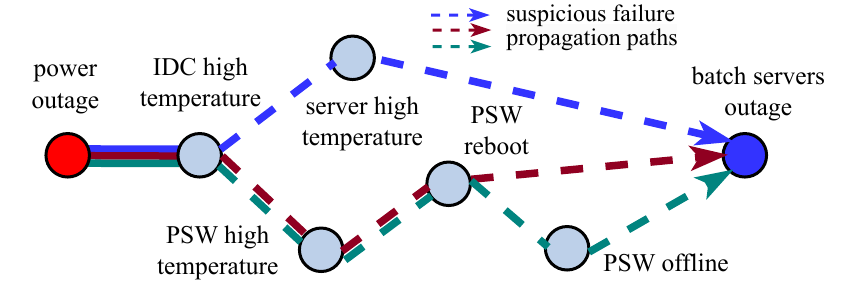}
  \caption{Analysis of real-world troubleshooting process}
  \label{fig:troubleshooting}
  \vspace{-0.5cm}
\end{figure}

\header{Analysis of Efficient Troubleshooting (RQ3)}.
Following an outage, the outage diagnosis system delivers diagnosis outcomes to
OSEs to facilitate rapid troubleshooting and system restoration.
Existing methods for failure diagnosis in microservice systems predominantly
employ root cause ranking or supervised learning approaches to predict the
top-$k$ root causes directly.
Nevertheless, these methods neglect the significant gap between diagnosis
outcomes and the troubleshooting process required in the real-world industrial
systems.
To illustrate the inadequacies of these methods in practical applications, we
present a troubleshooting analysis in \cref{fig:troubleshooting}.

In the practical troubleshooting process, to prevent the recurrence of
failures~\cite{li2022actionable}, OSEs systematically check all related
devices along the failure propagation path,
ensuring that all potential failures are repaired.
For example, in the case presented in \cref{fig:troubleshooting}, OSEs need to
investigate three suspicious paths.
In this process, OSEs discover that, in addition to the root cause (i.e., power outage),
aging of the PSW (Polymerize Switch) is another reason for this outage.
The aged PSW's tolerance for high temperatures decreased, leading to multiple
automatic reboots and causing the batch servers outage, ultimately.
Consequently, in addition to repairing the power components, replacing the PSW
is also necessary.
However, prior diagnosis methods focus solely on pinpointing the root cause
failure of outage, without providing insight into how the root cause failure
propagated step by step.
This limitation leads to the oversight of potential failures, posing a threat to
system stability.
Hence, {\em providing the interpretable diagnosis results that include both
root cause failure and failure propagation path is
necessary for efficient troubleshooting.}

\section{Batch Servers Outage Diagnosis}
\label{sec:problem}

In this section, we first give the related definitions, and then provide a
formal description of our problem.

\header{Anomaly, Failure, Event}.
We use anomaly, failure, and event to refer to abnormalities at various levels,
respectively.
An {\em anomaly} refers to an abnormal observation in device metrics, but does
not necessarily cause system failure.
A {\em failure} is a more serious malfunction that affects the system operation.
An {\em event} denotes an aggregation of similar failures or system changes that
occur nearby in time and on neighboring devices, typically sharing the same
triggering mechanism.

\header{Outage Snapshot}.
An outage snapshot refers to the collection of monitoring data related to the
outage.
Formally, let us set the outage occurrence time as the time origin, and use a
negative (positive) value to indicate the time amount before (after) the outage.
Then the outage snapshot is defined as $U \equiv \{\CA_{-L:T'}, \CI_{-T:T'},
\CC_{-T:T'}\}$, where $\CA$, $\CI$, and $\CC$ represent the collected alerts,
incidents, and changes, respectively; $L$, $T$, and $T'$ denote 
pre-set timestamps for data collection.

\subsection{Problem Formulation}
\begin{problem}[The Batch Servers Outage Diagnosis Problem]
  Given an outage case with outage snapshot $U$, we want to provide accurate and
  interpretable outage diagnostic results.
  Specifically, we have two sub-problems, i.e., failure detection
  and outage root cause analysis.
  \begin{itemize}
  \item Failure detection sub-problem takes the outage snapshot
    $U$ as input, detects all outage-related {\em events} $E$ in $U$ through a discriminator $\CF\colon U\mapsto E=\{e_1,\dots\}$.
  \item Outage root cause analysis sub-problem takes $E$ as input, locates the
    top-$k$ root cause set $S_U$ of the outage and infers the failure propagation
    path $p_U$ through a localizer $\CM\colon E\mapsto \{e_r, p_U\}$.
  \end{itemize}
\end{problem}

\section{Methodology}
\label{sec:method}

\begin{figure}[t] \centering
    \includegraphics[width=\linewidth]{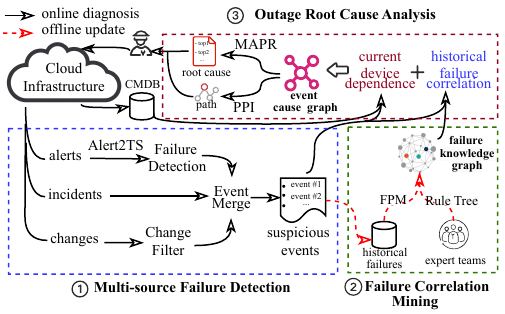}
    \caption{The overview of BSODiag}
    \label{fig:overview}
    \vspace{-0.6cm}
\end{figure}

The empirical observations in \cref{sec:empirical_analysis} reveal three
challenges in batch servers outage diagnosis problem: (i) low-quality failure
monitoring data, (ii) intricate failure correlations, and (iii) lack of
interpretability in root cause localization results.
In this section, we propose an unsupervised global diagnosis framework called
BSODiag.
\cref{fig:overview} provides an overview of BSODiag, illustrating its three core
modules, i.e., {\em multi-source failure detection}, {\em failure correlation
  mining}, and {\em outage root cause analysis}, which are designed to address
these respective challenges.
In what follows, we elaborate on each module in proposed BSODiag.

\subsection{Multi-source Failure Detection (MFD)}

When the monitoring platform detects a batch serves outage, BSODiag is
immediately activated.
The MFD module takes the outage snapshot $U$ as input, detects all suspicious
outage-related failures, and treats them as candidates for root cause analysis.
Our quality analysis in \cref{sec:empirical_analysis} indicates that
single-source coarse-grained monitoring data used by existing
methods~\cite{chen2019outage,wang2021groot} exhibits significant omissions
(mainly from incidents) or false positives (mainly from alerts) in
characterizing failures.
A direct solution to address this problem is to simultaneously collect and
analyze multi-source monitoring data~\cite{lee2023eadro,zhao2023robust}.
However, the inherent differences in data forms and attributes in multi-source
data hinder the direct application of this approach.
Therefore, in the MFD module, we first use modality-specific failure detection
components to detect genuine outage-related failures from different data sources
separately, then aggregate related failures and uniformly represent them as
events.
Considering that incidents have already been aggregated into high-risk events,
we only process alerts and changes during this stage.

\subsubsection{Alert2Event}
Alerts meticulously document all device-level anomalies, providing detailed
failure information.
However, we observe that due to the inherent triggering mechanism, alert
processing faces two significant challenges, i.e., false positives and alert
flooding.
False positive alerts are reports of abnormal device states, but actually not
related to the outage.
They are caused by the empirical thresholding based triggering mechanism
employed by the alert system.
On the other hand, alert flooding occurs from the repeated reporting of the same
failures, causing critical failures to be buried and difficult to identify.
Therefore, we design the Alert2Event sub-module to efficiently detect genuine
outage-related failures from the original alerts.
The core idea of Alert2Event is to compare the differences of alert patterns
(i.e., frequency and intensity of alerts) across different time periods before
and after an outage.
Intuitively, if a failure's alert pattern changes suddenly near the outage
occurrence time, it is likely to be related to the outage.
In contrast, if an alert pattern remains stable and consistent over time, it is
likely to be noise and unrelated to the outage.
Therefore, the Alert2Event sub-module consists of two components, namely {\em Alert2TS}
and {\em Failure Detection}, where Alert2TS converts raw alert data into
{\em alert sequences} to reflect the failure's alert pattern, and failure detection
identifies genuine outage-related failures from them.

\header{Alert2TS}.
We observe that a faulty device (i.e., one appearing in the outage snapshot) may
trigger alerts with different failure types and exhibit different
alert patterns.
Therefore, the Alert2TS component is designed to transform the raw alerts of
each faulty device into a multivariate time series, where each dimension of the
time series quantifies the alert pattern for a specific failure on that device.
Formally, for an alert set $\CA$ from the outage snapshot $U$, we first divide
$\CA$ into multiple alert subsets $\{\CA_1, \CA_2, \dots, \CA_S\}$ according to
their unique device serial number.
Each alert subset is defined as $\CA_s \equiv \{a_s^{k,t}\colon k\in\CK, t\in
[-L, T']\}$, recording all alerts that occurred on device $s$ in time range $t\in
[-L,T']$ in chronological order, where $a_s^{k,t}$ represents an alert that appeared
at time $t$ with failure type $k$, and $\CK$ represents all of the failure
types.
Next, we uniformly divide the timeline of outage snapshot into time slots of
length $\delta$, and aggregate the homogeneous alerts within each time slot.
Each alert subset $\CA_s$ is thus converted into an equidistant multivariate time
series $X_s\in\Real^{K\times N}$, where $K=|\CK|$ and $N = (L+T')/\delta$.
In the aggregation process, for a numerical alert, we extract the numerical
value $v_s^{k,t}$ from its description fields through regular expression
matching and set the alert intensity $r_s^{k,t} = v_s^{k,t}$; for non-numerical
alerts, we simply set $r_s^{k,t} = 1$ if failure type $k$ occurs at time slot
$t$ on device $s$.
Finally, each numerical point $x_s^{k,i}$ in the alert sequence $X_s$ is defined
as the cumulative alert intensity within the corresponding time slot, i.e.,
$x_s^{k,i} = \sum_{t\in l_i}r_s^{k,t}$, where $l_i$ represents time slot
$[(i-1)\delta-L, i\delta-L]$.

\header{Failure Detection}.
To accurately identify outage-related failures, the failure detection component
employs SPOT~\cite{ahmed2017detecting} algorithm for outlier detection in alert sequences.
SPOT is an unsupervised anomaly detection algorithm that can
efficiently identify outliers in a time series.
In our case, for an alert sequence $X_s$, SPOT considers the alert
sequence within the {\em initial window}, i.e., $t \in [-L,-T]$, as the normal
alert pattern to establish the extreme value boundaries.
It then dynamically updates these boundaries in the {\em diagnosis window},
i.e., $t\in[-T,T']$, and identifies outliers in it.
Outliers appearing in the same dimension are further consolidated into a single
failure $f$ along
with an attribute tuple \((sn, type\_failure, type\_device, start\_time,
end\_time)\), which describes the device serial number, failure type,
device type, start and end times of this failure.
Finally, each alert sequence $X_s$ is mapped to a failure set $\{f_1, f_2,
\dots, f_H\}$ that contains $H$ genuine failures discovered on
device $s$.

\subsubsection{Change Filter}

Different from alerts and incidents, 
changes record alterations in the system status.
The changes attended within a CIS can be classified into two
categories, i.e., {\em proactive changes} and {\em passive changes} 
according to the difference in trigger mechanism.
Proactive changes are initiated by OSEs or software engineers and include
actions such as hardware replacements, software fixes, or upgrades.
In contrast, passive changes are automatically triggered by external factors
such as automated virtual resource migrations.
Our empirical observations indicate that only proactive changes have the
potential to cause other failures.
Therefore, for the changes in the outage snapshots, we employ a rule-based whitelist to
remove irrelevant changes, retaining only the proactive changes for subsequent
outage investigation.

\subsubsection{Event Merge}

The failure detection component picks out the outage-related failures and
changes from multi-source monitoring data as candidates for further root cause
analysis.
However, these results are not immediately usable due to two issues, i.e., the
redundant records of the same failure across different monitoring sources and
the inconsistencies in the granularity of failure descriptions.
To address these issues, we employ an event merge component to integrate
failures from alerts, original incidents, and proactive changes.
Specifically, we segment the diagnosis window $[-T,T']$ into multiple time
windows of length $\eta$.
Within each time window, similar failures (those with the same failure type and
occurring on similar devices) are merged into an event, uniformly represented as
an attribute tuple \((sns, type\_failure, type\_device, start\_time, end\_time)\),
where $sns$ denotes the serial numbers of the failure devices. The start and
end times of an event correspond to its earliest and latest occurrence times.

\subsection{Failure Correlation Mining (FCM)}

After detecting all outage-related events $E$ in $U$, BSODiag is expected to
further identify the root cause of the outage.
However, accurately modeling the correlations between failures from a global
perspective is a challenging task.
Although these failures appear simultaneously in the outage snapshot, their
correlations may differ significantly.
For instance, given a failure pair $\langle f_a, f_b \rangle$ from $U$, if $f_b$
is caused by $f_a$, a strong causal correlation $f_a\rightarrow f_b$ exists
between them.
If $f_a$ and $f_b$ are simultaneously caused by another failure $f_c$, just a
concurrent correlation is present.
Conversely, if $f_a$ and $f_b$ merely coincidentally appear together, they are
unrelated.
In traditional industrial diagnosis scenarios, failure correlations are stored
as domain knowledge in the minds of different experts.
However, manual diagnosis is time-consuming and labor-intensive.
The FCM module aims to thoroughly explore failure information in historical
data, mining stable failure mechanisms to measure the correlations between
failures.

The basic idea of FCM module is to mine failure pairs with a high concurrent
frequency from historical data to represent the influence mechanisms between
failures.
For a failure pair $\langle f_a, f_b \rangle$, if we observe a high
conditional probability $P(f_b|f_a)$ in the historical data, it means that
there is likely a causal correlation $f_a\rightarrow f_b$.
These high-frequency failure pairs are further abstracted as failure knowledge,
serving the subsequent root cause diagnosis.

FCM module employs the classic association mining algorithm
Apriori~\cite{apriori2014imp} to mine high-frequency failure pairs from
historical data.
We gather historical failure data to construct an {\em event set} $\CT$.
The events in $\CT$ are then divided into multiple {\em event groups} $\CT =
\{\CT_1, \CT_2, \dots \}$, where each group $\CT_i = \{e_1, e_2, \dots \}$
shares the same spatio-temporal characteristics, indicating they occur on the
same day and within the same data center.
The event group is the maximum scope within which two failures can influence each
other.
It should be noted that when we focus only on the failure type attribute of
events, events and failures are equivalent.
Therefore, for convenience, we refer to $\langle e_a, e_b \rangle$ as a
failure pair.
For a failure pair $\langle e_a, e_b \rangle$, the metrics {\em support} and
{\em confidence} are used to measure the reliability of their correlation and
are calculated as:
\[
\vspace{-0.1cm}
\text{support}(\langle e_a, e_b \rangle)) = P(e_a, e_b) = \frac{num(\langle e_a, e_b \rangle)}{num(failure \ pairs)},
\]
\[
\text{confidence}(\langle e_a, e_b \rangle)) = P(e_b|e_a) = \frac{num(\langle e_a, e_b \rangle)}{num(e_a)}.
\]

We note that both high-frequency {\em causal} failure pairs and {\em concurrent} failure
pairs are discovered in this process, but only causal correlations are necessary
for root cause analysis.
Consequently, we further constrain the association mining using an expert-provided
failure rule tree, which depicts the
hierarchical relationships among failures from a macro perspective.
Only upper-level failures can trigger lower-level failures.
Therefore, we retain only high-frequency failure pairs that satisfy these
hierarchical relationships.

\cref{alg:apriori} gives the pseudo-code of the high-frequency failure pairs
mining (FPM) algorithm.
In Lines~\ref{ln:init_start}--\ref{ln:init_end}, high-frequency failures in the
historical data are initially filtered from the event set $\CT$.
Subsequently, in \cref{ln:pair}, candidate failure pairs are constructed, and
those that meet the failure hierarchical relationships (in \cref{ln:rule}) and
support thresholds (in \cref{ln:support}) are further picked out.
Finally, all high-frequency failure pairs set $\CQ_2$ and their confidence values are
output and stored in the failure knowledge graph $G_f$.

\begin{algorithm}[t]
  \caption{Failure Pairs Mining (FPM)\label{alg:apriori}}
  \KwIn{Event Set $\CT=\{ \CT_1,\CT_2,\dots\}$, Support Threshold $\alpha$}
  \KwOut{Frequent Failure Pairs Set $\CQ_2$}
  \tcp{frequent failures initialization}
  $\CQ_1\gets\emptyset$\; \label{ln:init_start}
  \For{each event $e \in \CT$}{
    \lIf{$e \in \CQ_1$}{$e.count\gets e.count + 1$}
    \Else{$\CQ_1\gets e$\;
          $e.count\gets 0$\;}
  $\CQ_1\gets\{e\in\CQ_1\mid e.count\geq |\CQ_1|\cdot\alpha\}$\;\label{ln:init_end}
  }
  \tcp{frequent failure pair mining}
  $\CQ_2\gets\emptyset$\;
  \For{$i\gets 1,\ldots,|\CQ_1|$}{
    \For{$j\gets i,\ldots,|\CQ_1|$}{
        $p_{ij}\gets \langle e_i, e_j \rangle$\; \label{ln:pair}
        \If{$e_i.level > e_j.level$}{\label{ln:rule}
            $\CQ_2\gets p_{ij}$\;
            \For{each failures group $\CT_l\in\CT$}{
              \lIf{$p_{ij}\in\CT_l$}{$p_{ij}.count\gets p_{ij}.count+1$}
            }
        }
    }
  }
  $\CQ_2\gets\{p_{ij}\in\CQ_2\mid p_{ij}.count\geq |\CQ_2|\cdot\alpha\}$\;
  \label{ln:support}
\end{algorithm}

\subsection{Outage Root Cause Analysis (ORCA)}

In a CIS, the interconnections between devices are intricate, involving
correlations such as physical connections, resource sharing, and service
dependencies.
Failures propagate gradually along these inter-device dependencies, ultimately
leading to a batch servers outage.
The ORCA module aims to comprehensively analyze the correlations between
outage-related failures, accurately locate the root cause of outage, and infer
the failure propagation path, providing OSEs with interpretable diagnosis
results.
ORCA mainly consists of three components, i.e., event cause graph construction,
outage root cause location, and failure propagation path inference.
The detailed descriptions of each component are as follows.

\subsubsection{Event Cause Graph Construction}

To provide a comprehensive view for failure analysis, we construct an event
cause graph $G_e$ for each outage case to describe the correlations between
failures.
$G_e$ is a directed graph where each node $e_i$ represents an outage-related
event in the failure snapshot $U$.
The direction of the edges indicates the triggering correlation between
failures, and the edge weight $\omega_{ij}$ represents the correlation strength
of a failure pair $\langle e_i, e_j \rangle$.

The failure knowledge graph $G_f$ models failure correlations within historical
data.
Intuitively, the failure knowledge in $G_f$ can be used to construct the event
cause graph $G_e$.
For example, for two events $e_i$ and $e_j$, a naive strategy to measure
their correlations is to search the failure pair $\langle e_i, e_j \rangle$.
If the edge $e_i \rightarrow e_j$ exists in $G_f$, we set $w_{ij} = 1$;
otherwise, $w_{ij} = 0$.
However, this approach has two significant drawbacks.
First, such binary edges cannot accurately model the varying propagation
capabilities between different failures.
Second, it relies entirely on historical information and cannot apply
to new failures that have never appeared before.

To address these issues, we further propose a spatio-temporal failure
correlation strategy that analyzes failure information from both historical and
current perspectives simultaneously.
Specifically, we obtain historical failure correlations from the constructed
fault knowledge graph $G_f$, using the confidence of failure pair
$\langle e_i, e_j \rangle$, denoted as $e_{ij}.conf$, as the measure of
reliability of failure correlation $e_i \rightarrow e_j$.
Additionally, we utilize a configuration management database (CMDB) that records
detailed system configurations and architectural information to assess the
current physical connectivity of failures.
For an event $e_i$, we let $e_i.sn$ denote the devices affected by $e_i$, and
let $(e_i.out).sn$ denote all peripheral devices connected to $e_i.sn$.
We define the connectivity strength between two events $e_i$, $e_j$ as
$dist(e_i, e_j)=|(e_i.out).sn \cap e_j.sn|/|e_i.sn|$.
Particularly, when $e_i$ and $e_j$ are the same type of device, the connectivity
strength is simplified to $dist(e_i, e_j)=|e_i.sn \cap e_j.sn|/|e_i.sn|$.
Finally, we combine the spatio-temporal information of the failures to calculate
the causal strength of each failure pair $\langle e_i, e_j \rangle$ as 
$w_{ij} = \exp(p_{ij}.conf) \cdot dist(e_i, e_j)$.

\subsubsection{Outage Root Cause Location}

The event cause graph preliminarily reveals the correlations between
failures from a global perspective.
In real industrial diagnostics, expert teams jointly analyze multiple factors
such as the occurrence time, severity, and propagation capability of failures,
gradually deducing the propagation process between failures to identify the root
cause.
Inspired by this, we propose a customized multi-attribute event graph random
walk algorithm (MAPR) to locate the outage root cause.

MAPR first introduces the time priority and impact on the outage of each event
into the node attribute initialization, defining node personalization score as
$u_i = exp(-t)\cdot dist(e_i, e_o)$,
where $e_o$ is the outage node.
Subsequently, it uses the failure correlation $w_{ij}$ as the failure transition
probability of $e_i\rightarrow e_j$, and performs multiple walks on the event
graph, continuously updating the node personalization score $u_i$.
It is worth noting that to prevent the walk from getting stuck in outage nodes,
we add an incoming edge $e_o\rightarrow e_i$ to each outage-related node $e_i$
and set the edge weight $w_{oi} = \frac{1}{|E|-1}$.
After $L$ iterations, the personalization score of each node $e_i$ converges to
a constant $\bar{u}_i$.
We then remove the outage nodes and determine the top-$k$ root cause nodes set
$S_U$ by ranking the personalization scores of the remaining nodes.

\subsubsection{Failure Propagation Path Inference}

Although we have identified root cause node $e_r = S_U[1]$, i.e., top-$1$ node in
$S_U$, the propagation path
from $e_r$ to $e_o$ remains unclear.
The empirical analysis in \cref{sec:empirical_analysis} indicates that providing
interpretable failure propagation paths is crucial for enhancing the
troubleshooting efficiency of OSEs.
To this end, we further propose a propagation probability based failure
propagation path inference method (PPI).
The basic principle of PPI is to select the path with the highest cumulative
propagation probability from the root cause node to the outage node on the event
causality graph.
Specifically, we first search all connected paths $\CP =\{p_1, p_2, \dots\}$
from $e_r$ to $e_o$.
Then, we calculate the cumulative propagation probability of each path and
select the one with the highest probability:
\[
\vspace{-0.2cm}
  p_U
  =\arg\max_{p_i\in\CP}\text{TransPr}(p_i)
  =\arg\max_{p_i\in\CP}\prod_{j\in |p_i|}\bar{u}_j,
\]
where $|p_i|$ indicates the number of nodes in each path $p_i$.

\section{Experiments}
\label{sec:experiments}

In this section, we perform experiments on real-world diagnosis data to evaluate
the performance of the proposed BSODiag method.

\subsection{Experimental Setting}

\subsubsection{Dataset Collection}

We built a large-scale testing platform in Alibaba CIS and collected
all monitoring data in this testing platform from January 2022 to December 2023.
This dataset includes alerts, incidents, and changes, encompassing $95$ batch
server outages, each affecting dozens to hundreds of servers.
We divided the collected monitoring data into four distinct datasets.
The initial dataset, $\CD_{init}$, comprises all monitoring
data from 2022 and serves to initialize BSODiag.
The remaining three datasets, i.e., $\CD_{idc}$, $\CD_{net}$, and $\CD_{all}$,
contain monitoring data from 2023 and are used to validate the diagnostic performance.
Specifically, $\CD_{all}$ includes data from all outage cases in 2023, while $\CD_{idc}$
and $\CD_{net}$ are subsets focusing on outages caused by IDC failures and network
failures, respectively.
The detailed statistics of these datasets are presented in \cref{tab:dataset}.

\begin{table}[htp]
  \centering
  \caption{Datasets statistics\label{tab:dataset}}
  \begin{tabular}{c|ccccc}
    \hline
    \hline
    Dataset & \#Incident & \#Alert & \#Change
    & \makecell{\#Failure\\Types} & \makecell{\#Outage\\ Cases} \\
    \hline
    $\CD_{init}$  & $19,020$ & $879,870$   &$3,255$  & $62$ & $27$    \\
    $\CD_{idc}$  & $173$ & $256,212$   &$1,657$  & $31$ & $19$    \\
    $\CD_{net}$  & $478$ & $774,638$   &$5,091$  & $44$ & $47$    \\
    $\CD_{all}$  & $665$    & $1,032,851$ & $7,644$ & $56$ & $68$   \\
    \hline
  \end{tabular}
\end{table}

In the experiment, based on the diagnostic experience, we set the
initial window for outage snapshots to $[-4, -2]$, indicating the range from $4$
hours to $2$ hours before the outage.
The diagnosis window is set to $[-2, 0.25]$, spanning from $2$ hours before the
outage to $15$ minutes after the outage.
Additionally, the time slot length $\delta$ and the event merge time window
$\eta$ are set to 1 minute and 5 minutes, respectively.
Support threshold $\alpha$ in \cref{alg:apriori} is set to $0.001$, and the number of iterations $L$ for MAPR is set to $100$.

\subsubsection{Compared Baselines}

We compare BSODiag with three rule based methods, two machine learning
(ML) based methods, and two failure graph based methods.
Rule based and ML based methods are prevalent in industrial settings for
their straightforward implementation, while failure graph based methods are
state-of-the-art approaches in current microservices outage diagnosis.

\noindent\textbullet\ \emph{Rule based methods}.
  We employ two widely used root cause diagnosis strategies:
  \textbf{Hierarchy-First} and \textbf{Time-First}.
  The hierarchy-first strategy~\cite{liu2021microhecl} determines the hierarchy
  of suspicious failures based on a failure rule tree and considers the failure
  with the highest hierarchy as the root cause.
  In contrast, the time-first strategy~\cite{meng2020localizing} identifies the
  earliest occurring failure as the root cause.
  In addition, we adopt a \textbf{Random Selection} strategy to simulate the
  diagnosis process without any domain knowledge.
  To improve diagnosis accuracy, we apply these strategies on the output of MFD
  module.
  
\noindent\textbullet\ \emph{Machine learning based methods}.
  These methods model the root cause location problem as a supervised
  classification problem, and address it using machine learning techniques.
  We implement this approach by combining the MFD module with a ML-based
  classifier.
  Each event output by the MFD module is initially represented as a
  multi-attribute vector encompassing failure time, failure hierarchy, and its
  connectivity strength with outage event.
  Subsequently, an \textbf{SVM} or \textbf{Random Forest} classifier is utilized
  to identify the root cause.
  
\noindent\textbullet\ \emph{Failure graph based methods}.
  \textbf{AirAlert}\cite{chen2019outage} and \textbf{COT}~\cite{wang2021outage}
  are state-of-the-art methods in microservices outage diagnosing problem.
  AirAlert leverages the Fast Causal Inference algorithm~\cite{colombo2012fci}
  to discern causal relationships between alert sequences and outages,
  constructing a failure graph, then using an XGBoost model to pinpoint the root
  cause.
  COT constructs an incident correlation graph using historical incident data,
  models the connections between failure based on this, and uses an SVM
  classifier to determine the root cause.

\subsubsection{Evaluation Metrics}

In outage diagnosis, BSODiag performs two tasks, i.e., root cause localization
(RCL) and failure propagation path inference (PPI).
For RCL task, we use two metrics PR@k and MAP to assess the performance of
BSODiag and other baselines.
We use $\mathcal{U}$ to denote all outage cases in $\CD_{diag}$, where each
outage case $U \in \mathcal{U}$ is represented by the corresponding outage
snapshot.
PR@K is defined as $ PR@k = \frac{1}{|\mathcal{U}|}\sum_{U \in \mathcal{U}}\mathbf{1}[r_U \in S_U[1:k]]$,
where $r_U$ is the true root cause of outage case $U$, and $S_U[1:k]$ represents
the top-$k$ root causes predicted by the diagnosis method.
Considering that accurately locating the top-$1$ root cause is challenging, we set
$k= \{1,2,3\}$.
MAP further evaluates the average performance for different $k$, calculated as
$MAP = \frac{1}{k}\sum_{1\leq i \leq k}PR@i$.

For PPI task, we define path coverage rate (PCR) to evaluate the explainability of the
inferred propagation paths. PCR is defined as
$PCR = \frac{1}{|\mathcal{U}|}\sum_{U \in \mathcal{U}}\frac{|p_U \cap \hat{p}_U|}{|p_U|}$,
where $p_U$ and $\hat{p}_U$ denote the predicted failure propagation path and
its ground truth, respectively.
$|p_U \cap \hat{p}_U|$ represents the longest common subpath of $p_U$ and
$\hat{p}_U$.
For all three metrics, larger values indicate better performance.

\subsection{Performance}

\begin{table*}[t]
  \centering
  \caption{Comparison of different methods for RCL task
    \label{tab:rcl}}
  \begin{tabular}{c|cccc|cccc|cccc}
    \hline
    \hline
    \multirow{2}{*}{Methods} & \multicolumn{4}{c|}{$\CD_{idc}$} & \multicolumn{4}{c|}{$\CD_{net}$} 
    & \multicolumn{4}{c}{$\CD_{all}$}  \\ \cline{2-13} 
    
    & \multicolumn{1}{c|}{PR@1} & \multicolumn{1}{c|}{PR@2} & \multicolumn{1}{c|}{PR@3} & MAP & \multicolumn{1}{c|}{PR@1} & \multicolumn{1}{c|}{PR@2} & \multicolumn{1}{c|}{PR@3} & MAP & \multicolumn{1}{c|}{PR@1} & \multicolumn{1}{c|}{PR@2} & \multicolumn{1}{c|}{PR@3} & MAP \\ \hline
    Random Selection      &14.3\% & 31.2\%& 42.6\%&39.4\% & 8.7\%&23.8\% & 36.4\%&23.0\% & 10.4\%   & 28.8\%   & 39.6\%   & 25.4\%   \\
    Hierarchy-First          & 12.6\%&27.5\% &46.3\% &28.8\% &13.9\% & 37.8\%& 66.4\%& 39.4\%& 12.5\%   & 35.4\%   & 62.5\%   & 36.8\%   \\
    Time-First            & 22.5\%&42.6\% &53.8\% & 39.6\%&41.2\% &56.0\% & 73.7\%& 57.0\%& 35.0\%   & 52.0\%   & 70.1\%   & 52.4\%   \\
    \hline
    SVM                  & 32.0\% & 44.6\%& 62.5\%& 46.4\%& 27.4\%& 44.2\%&65.3\% &45.6\% &  27.8\%   & 43.9\%  & 66.1\%  &  45.9\%   \\
    Random Forest         & 39.2\%&58.8\% & 72.0\%& 56.7\%& 41.4\%& 57.9\%& 71.4\%& 56.9\%& 42.6\%   & 58.7\%  & 74.3\%  &  58.5\%    \\
    \hline
    AirAlert             & 18.5\%& 30.9\%& 41.0\%& 30.1\%&28.0\% &43.2\%   & 53.6\%& 41.6\%& 24.5\%   & 38.7\% & 48.8\%  &  37.3\%  \\
    COT                  & 46.3\%& 66.0\%& 82.7\%& 65.0\%& 40.8\%& 57.5\% & 72.2\%&56.8\% & 44.9\%   & 62.4\%  & 77.3\%  &  61.5\%  \\
    \hline
    BSODiag (\textbf{ours})  & \textbf{56.1\%} & \textbf{72.9\%} & \textbf{88.2\%} & \textbf{72.4\%}& \textbf{52.4\%}& \textbf{70.7\%}& \textbf{86.7\%} &\textbf{69.9\%} & \textbf{54.2\%} & \textbf{70.8\%} & \textbf{87.5\%} & \textbf{70.8\%} \\
    \hline
  \end{tabular}
  \vspace{-0.3cm}
\end{table*}

We first examine the performance of BSODiag and other baselines on the RCL task.
As shown in \cref{tab:rcl}, BSODiag significantly outperforms all other
baselines across all metrics.
In dataset $\CD_{all}$, compared to the most competitive baseline, i.e., COT,
BSODiag improved by
$9.3\%$, $8.4\%$, $10.2\%$, and $9.3\%$ on the PR@1, PR@2, PR@3, and MAP,
respectively.
Similar advantages are also observed in $\CD_{idc}$ and $\CD_{net}$ datasets.
This indicates that BSODiag can more accurately locate the root cause of
outages.
Additionally, we observe that although rule-based methods are straightforward,
their performance is suboptimal because they only consider the failures
attributes superficially and cannot fully perceive the correlation between failures.
Meanwhile, for ML-based methods, the rarity of outage cases makes it difficult
to sufficiently optimize the model.
This further illustrates that our proposed unsupervised diagnosis strategy
based on the event cause graph is more suitable for the outage diagnosis problem.

Comparing the performance of all methods on $\CD_{idc}$ and $\CD_{net}$,
we observe that rule based and failure graph based baselines exhibit obvious
performance fluctuations. Notably, rule based methods demonstrate a performance
difference of approximately 20\% on $\CD_{net}$ compared to $\CD_{idc}$.
This phenomenon stems from inherent differences in failure propagation mechanism 
across various domains.
IDC failures typically span multiple domains, whereas network failures present
a more straightforward hierarchical structure, facilitating root cause localization.
Although ML-based methods show consistent performance across different domains,
their root cause diagnosis capabilities remain suboptimal due to their inability
to model the intrinsic correlations between failures.
In contrast, our method,
i.e., BSODiag, comprehensively models failure correlations by incorporating
historical knowledge and current device dependencies, demonstrating consistent
superiority across various domains.

\begin{table}[htp]
  \centering
  \caption{Comparison of different methods for PPI task.\label{tab:fpi}}
  \begin{tabular}{c|c|c|c|c}
    \hline
    \hline
\multirow{2}{*}{Dataset} & \multicolumn{4}{c}{Methods} \\ \cline{2-5} 
                         & \multicolumn{1}{c|}{DPS} & \multicolumn{1}{c|}{SPS} & \multicolumn{1}{c|}{FHM} &  BSODiag(\textbf{ours}) \\ \hline
                         $\CD_{idc}$& \multicolumn{1}{c}{38.0\%} & \multicolumn{1}{c}{35.2\%} & \multicolumn{1}{c}{41.8\%} & \textbf{45.6\%} \\ 
                         $\CD_{net}$& \multicolumn{1}{c}{41.6\%} & \multicolumn{1}{c}{32.4\%} & \multicolumn{1}{c}{43.3\%} &  \textbf{46.8\%}\\ 
                         $\CD_{all}$& \multicolumn{1}{c}{40.2\%} & \multicolumn{1}{c}{ 33.8\%} & \multicolumn{1}{c}{42.6\%} &  \textbf{46.3\%} \\ 
    \hline
  \end{tabular}
  \vspace{-0.2cm}
\end{table}

For the PPI task, due to the absence of directly comparable baselines, we compare
BSODiag with several straightforward path inferring strategies, i.e.
\textbf{deepest path search} (DPS), \textbf{shortest path search} (SPS), and
\textbf{Failure Hierarchy Matching} (FHM).
DPS and SPS, respectively, search for the longest and shortest path between the
top-1 root cause node $e_r$ and outage node $e_o$ as the failure propagation
path.
The failure hierarchy matching strategy strictly follows the
``upstream-downstream'' propagation policy described in the failure rule tree to
infer failure propagation path.
The results, exhibited in \cref{tab:fpi}, indicate that BSODiag achieves
$46.3\%$ PCR, showing an improvement of $6.1\%$, $12.5\%$, and
$3.7\%$ compared to the other baselines, respectively.

\subsection{Online Evolution}
In BSODiag, the failure knowledge graph, constructed from historical data, is
essential for accurately measuring failure correlations.
The scale of historical data directly determines the quality of the constructed
failure knowledge graph, thereby affecting the diagnosis performance of BSODiag.
In actual online deployment, as more failure data are collected, we can
continuously update BSODiag to optimize its performance.

\begin{figure}[htp]
  \centering
  \vspace{-0.5cm}
  \subfloat[MAP]{\includegraphics[width=0.5\linewidth]{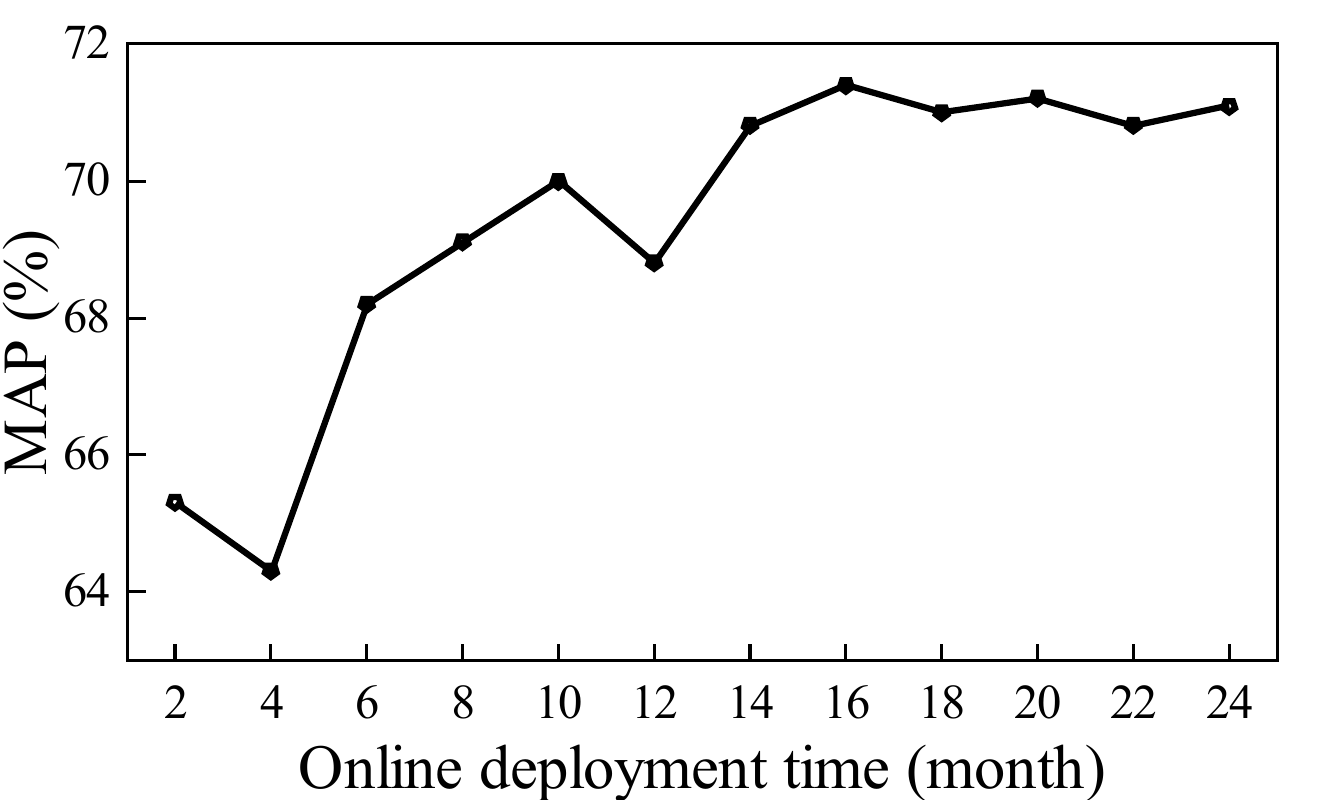}}
  \subfloat[PCR]{\includegraphics[width=0.5\linewidth]{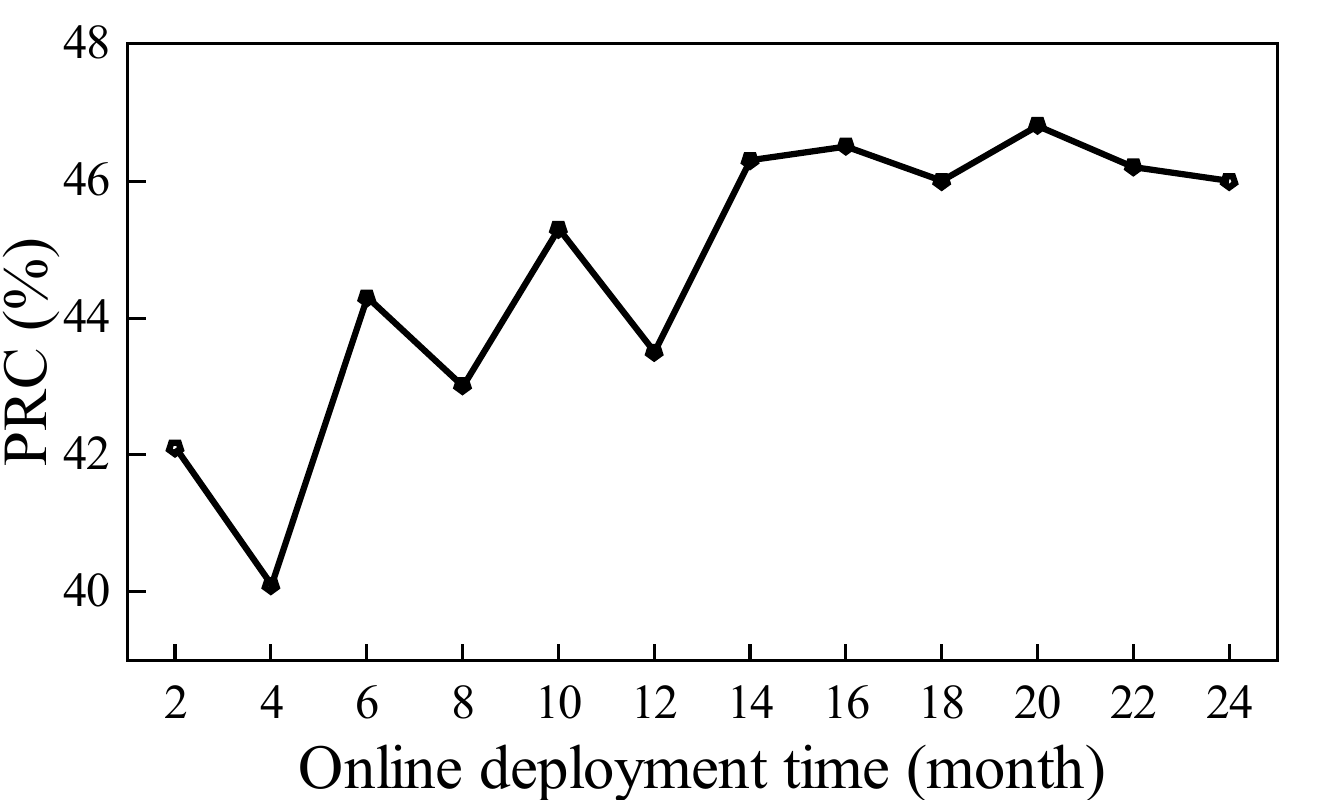}}
  \caption{The online deployment performance of BSODiag.
    \label{fig:deployment}}
\vspace{-0.3cm}
\end{figure}

Figure~\ref{fig:deployment} illustrates the online evolution of BSODiag's
performance over two years.
Despite minor fluctuations in root cause diagnosis performance due to the
deployment of differentiated services in real cloud systems, BSODiag's overall
performance exhibits an initial increase followed by stabilization.
In the early stages, when failure data are sparse, BSODiag's diagnostic accuracy
is limited due to insufficient fault knowledge. As more data accumulates,
the quality of the failure knowledge graph gradually improves, resulting
in enhanced diagnostic performance. In the later stages, the failure knowledge
graph reaches a quality plateau, leading to a steady state in BSODiag's performance.

\subsection{Ablation Study}

\begin{figure}[t]
    \centering
    \begin{minipage}[t]{0.49\linewidth}
        \centering
        \includegraphics[width=1.0\linewidth]{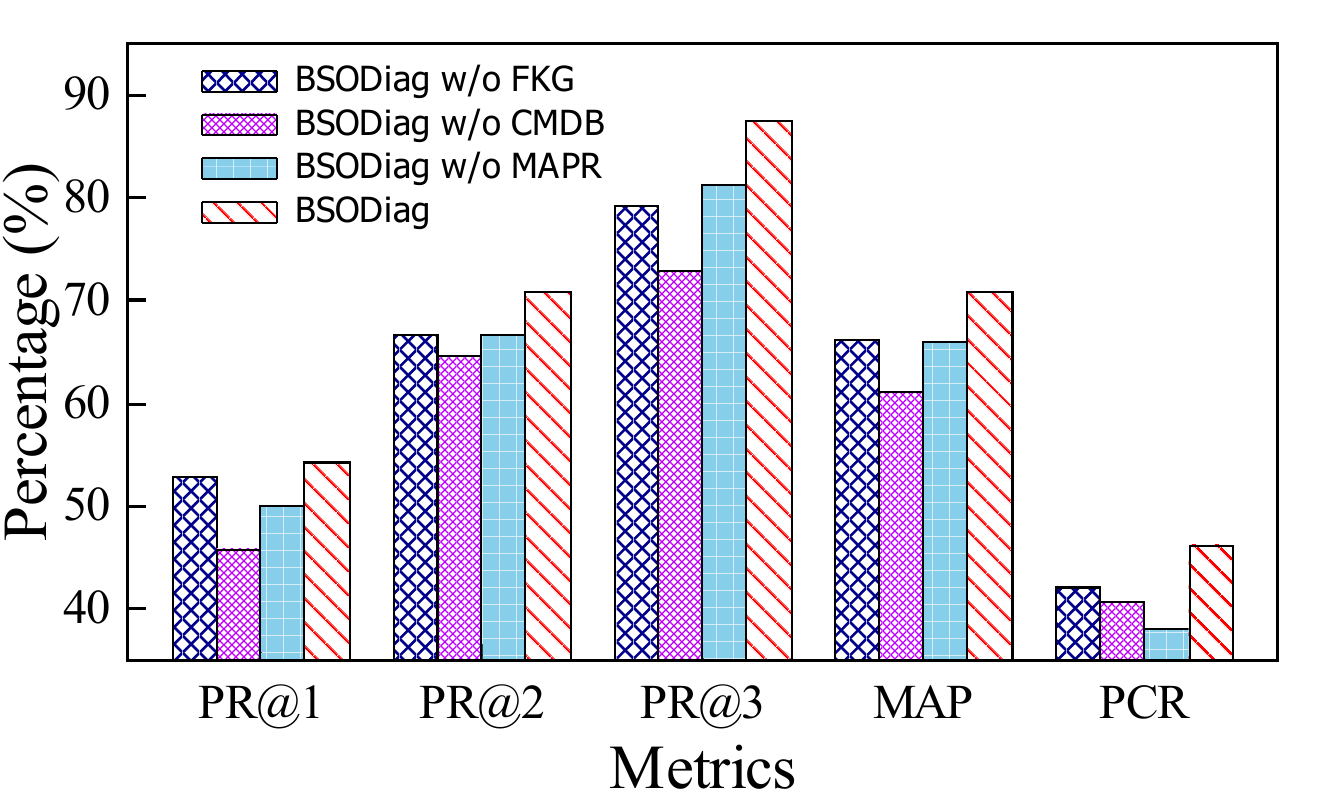}
        \caption{Ablation study}
        \label{fig:ablation}
    \end{minipage}
    \begin{minipage}[t]{0.49\linewidth}
        \centering
        \includegraphics[width=0.95\linewidth]{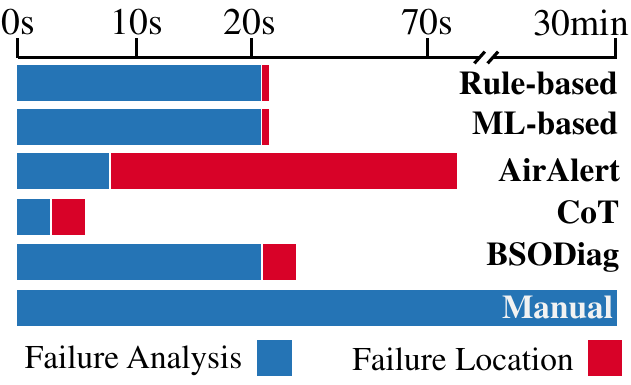}
        \caption{Time Consumption}
        \label{fig:rca_time}
    \end{minipage}
    \vspace{-0.5cm}
 \end{figure}

In this section, we conduct an ablation study on $\CD_{all}$ dataset to provide a better
understanding of the proposed BSODiag.
In the experiment, we remove each of the three core modules separately: the
failure knowledge graph, the configuration management database, and the event
cause graph multi-attribute random walk.
These are denoted as ``BSODiag w/o FKG'', ``BSODiag w/o CMDB'', and ``BSODiag
w/o MAPR'', respectively.
The results are shown in \cref{fig:ablation}.

We observe that when the corresponding components are removed, the performance
of BSODiag on both RCL and PPI tasks significantly declines.
This indicates that (i) the proposed spatio-temporal failure correlation analysis
strategy can provide more accurate failure correlation modeling; and (ii) the
proposed global analysis perspective can better understand the propagation process
of failures, thereby improving the ability to locate root causes and infer
failure propagation path.

\subsection{Efficiency Analysis}

To further evaluate the diagnosis efficiency of BSODiag, we conduct an
efficiency analysis in this section.
\cref{fig:rca_time} compares the root cause diagnosis efficiency of BSODiag with
all baselines and expert maintenance technicians in the Alibaba on $\CD_{all}$
dataset. The total time consumption is divided
into two stages, i.e., failure analysis and failure localization.

The results demonstrate that BSODiag achieves a single diagnosis of an outage case
in $24.5$ seconds, marking a substantial improvement over the traditional manual
diagnosis, which typically requires an average of $30$ minutes.
Additionally, we note that the most time-consuming part of the diagnosis process
is the MFD module, as it involves complex failure analysis operations including both 
failure detection and failure merging. Similarly, rule-based and ML-based methods
primarily consume time
during the failure analysis process. In contrast, AirAlert's time consumption is
mainly in the failure localization phase, as it necessitates the computation of
pairwise correlations among alert sequences, a process that demands significant
computational resources. 
While COT exhibits the highest diagnostic efficiency by processing only incident
data, our findings in~\cref{tab:rcl} suggest that this approach may overlook certain
outage-related failures, potentially compromising the accuracy of root cause localization.

\section{Related Work}
\label{sec:related_work}

\header{Failure Detection}.
Failure detection~\cite{ma2020automap,sun2023efficient} aims to promptly identify
anomalies using monitoring data
generated by hardware devices or service nodes.
Depending on the type of data required, existing methods can be categorized as
metric based, log based, trace based, and multi-modal based methods.

{\em Metric based} methods identify failures by modeling both the normal and abnormal
sequential patterns in metrics using traditional statistical methods~\cite{siffer2017SPOT}
or end-to-end learnable neural networks~\cite{zhang2022metric}.
{\em Log based} methods transform logs into keyword index
sequences~\cite{du2017deeplog,jia2023robustlog}
or natural language~\cite{meng2019loganomaly}, and detect failures by combining
the sequential and semantical information of logs.
{\em Trace based} methods are mainly used for detecting invocation anomalies in
microservice systems.
Such methods typically use historical traces to model the normal response pattern,
and then during the detection phase, traces that deviate from the normal
pattern are identified as anomalies~\cite{liu2020trace,xie2023trace}.
Multi-modal based methods jointly model metrics, logs, and traces, mining
failure patterns from multiple data modalities and reducing the rate of missed
failure detection~\cite{zhao2023robust,lee2023eadro}.

\header{Root Cause Analysis}.
Due to the complex correlations between devices or services in cloud service
systems, a single failure often triggers multiple related failures.
Root cause analysis (RCA)~\cite{ikram2022root,ma2021servicerank,chen2024automatic} aims to
locate the root cause failures~\cite{soldani2022anomaly}.
The existing RCA methods can be divided into search based,
learning based, and causality graph based methods.

Traditional {\em search based} RCA methods~\cite{ahmed2017detecting,sun2018hotspot}
identify root causes by searching for attribute
combinations with the highest commonality in failure. For example, HotSpot~\cite{sun2018hotspot}
uses a potential score to measure the correlation of different attribute combinations
within failures and employs a heuristic strategy to reduce the search space of
attribute combinations.
These methods are the easiest to deploy. However, it should be noted that they
are only applicable to scenarios with discrete failure attributes due to their
reliance on restricted attribute space.

{\em Learning based} methods~\cite{li2022actionable,lee2023eadro} typically model RCA
as a supervised graph node
classification problem.
These methods first construct a component dependency graph, where each node represents
a device or service node, and the edges represent the inherent connections between
corresponding components.
Graph neural networks are then used to aggregate failure
information among different nodes. 
For example, Eadro~\cite{lee2023eadro} first builds a node dependency graph
through component interaction information and then uses a GAT-based status learning approach
to accomplish failure
detection and root cause localization tasks, simultaneously.
Due to the requirement of a large amount of labeled data,
these methods are difficult to deploy in real industrial environments.

{\em Causality graph based} RCA methods~\cite{wang2021groot,chen2021graph} have
been extensively studied in recent years.
These methods typically contain two core steps, i.e., failure causality reasoning and
root cause
localization~\cite{wu2020microrca,meng2020localizing,li2021localization,wang2021groot}.
The failure causality reasoning constructs a failure causality graph that
directly reflects the failure propagation process by mining the correlation
between failures through their own information.
For instance, MicroCause~\cite{meng2020localizing} proposes a path condition
time series algorithm to learn the causal relationships between failures from
the metrics of microservice nodes.
COT~\cite{wang2021outage} obtains the correlations between failures from
historical incidents and constructs a failure causality graph based on this.
Subsequently, on the failure causality graph, these methods
use random-walk based or search based approaches to locate the root cause.

\section{Conclusion}
\label{sec:conclusion}

The failure diagnosis capability of cloud infrastructure systems is crucial for
maintaining system availability.
In this paper, we propose an unsupervised lightweight diagnosis framework BSODiag
to address the batch servers outage diagnosis problem. 
BSODiag comprehensively detects system failures by fully utilizing multi-source failure
monitoring data, including alerts, incidents, and changes.
Subsequently, failure spatio-temporal information reflected in historical
data and current devices is used to measure the correlation between failures. 
Finally, BSODiag analyzes the failure propagation process from a global perspective,
accurately locates the root causes of the outage, and infers failure propagation paths.
Experiments on real industrial data demonstrate that BSODiag significantly
outperforms existing baseline methods.

\section*{Acknowledgements}
\label{sec:acknowledgements}

We are grateful to anonymous reviewers for their constructive comments
to improve this paper. 
This work was supported in part by Alibaba Innovative Research (AIR) Program
and the National Natural Science Foundation of China (62272372, 61902305, U22B2019).

\clearpage

\bibliographystyle{IEEEtran}
\bibliography{IEEEabrv,ref}

\end{document}